\documentclass[12pt]{article}
\usepackage{graphicx}
\usepackage{amsfonts}
\usepackage{amssymb}
\usepackage{epsf}

\newcommand{\be}{\begin{equation}}
\newcommand{\ee}{\end{equation}}
\newcommand{\bea}{\begin{eqnarray}}
\newcommand{\eea}{\end{eqnarray}}

\newcommand{\nn}{\nonumber\\}

\newcommand{\vp}{\vec\phi}
\newcommand{\vpc}{\vec\phi_c}

\begin{document}

\hfill{KCL-PH-TH/2012-41}

\vspace{1cm}

\begin{center}

{\bf {\Large Maxwell Construction for Scalar Field Theories with Spontaneous 
Symmetry Breaking}}

\vspace{1cm}

{\bf J. Alexandre}$^a$\footnote{jean.alexandre@kcl.ac.uk} and {\bf A. Tsapalis}$^{b,c}$ \footnote{tsapalis@snd.edu.gr}\\

\vspace{0.5cm}

{\small {\it a} ~King's College London, Department of Physics, WC2R 2LS, UK}\\
{\small {\it b}~ Hellenic Naval Academy, Hatzikyriakou Avenue, Pireaus 185 39, Greece}\\
{\small {\it c}~ Department of Physics, National Technical University of
Athens \\ Zografou Campus, 157 80 Athens, Greece}

\vspace{1cm}

{\bf Abstract}

\end{center}

Using a non-perturbative approximation for the partition function of a complex scalar model, which features spontaneous symmetry breaking, 
we explicitly derive the flattening of the effective potential in the region limited by the minima of the bare potential. 
This flattening  
occurs in the limit of infinite volume, and is a consequence of the summation over the continuous set of saddle points which dominate the 
partition function. We also prove the convexity of the effective potential and 
generalize the Maxwell Construction for scalar theories with $O(N)$ symmetry.
Finally, we discuss why the flattening of the effective potential 
cannot occur in the Abelian Higgs theory.

\section{Introduction}

The convexity of the effective potential for a scalar theory has been known for a long time \cite{Symanzik},
and is a consequence of its definition in terms of a Legendre transform \cite{Haymaker}.
In the situation where the bare potential features spontaneous symmetry breaking (SSB), convexity
is achieved non-perturbatively, and cannot be obtained by a naive loop expansion. 
The effective potential becomes flat 
between the two minima of the bare potential, as a consequence of the competition of the two non-trivial saddle points \cite{Fujimoto}.
By analogy with the Maxwell construction for a Van de Waals fluid, the 
corresponding flattening can be understood from the so-called spinodal instability:
no restoration force suppresses fluctuations of the vacuum for the quantized system, which is a superposition of the two bare vacua \cite{Miransky}. 
A PhD thesis has been written on the topic \cite{Kessel}, where many aspects are studied, and a detailed literature
review is given.

We note that the equivalence between the effective potential defined via the Legendre transform and the Wilsonian effective potential 
is valid in the limit of infinite volume only \cite{Wipf}, as the Wilsonian effective potential is not necessarily convex at finite volume. 
Lattice simulations of scalar SSB models  in \cite{Wipf} show the gradual flattening for the
Wilsonian effective potential (the ``constrained effective potential'') as the number of lattice sites increases. On the other hand, as 
shown in the present work, the effective potential obtained by the Legendre transform is convex for finite volume too.

A linear effective potential was explicitly derived in \cite{alex}, for a real scalar field in a SSB bare potential. 
Using a simple approximation, it was shown that the Maxwell construction arises from the dominant contributions of both saddle points in the 
partition function. 
These dominant contribution are homogeneous and do not take into account the kink solution, which is stable in 1+1 dimensions only
if no other field is present 
\cite{Jackiw}. We consider here two space dimensions at least (a recent work on the quantization of the 1+1 
dimensional kink can be found in \cite{Rajantie}).

Quantum corrections to the (Legendre) effective potential are calculable in series of
$\hbar$ via functional techniques \cite{Jackiw2} as fixed-loop diagrammatic 
series.  
One- and two-loop corrections, computed in \cite{Jackiw2}
for the $O(N)$ theory, do not suffice to restore convexity for a tree-level
symmetry breaking potential. A large-N analysis,
performed at the leading $1/N$ order in \cite{Jackiw3}, leads to a 
real, convex 
potential in four dimensions albeit for a restricted range of field amplitudes,
raising thus questions to the consistency of the scheme in that order.
It is interesting though, that in lower dimensions the same approximation 
leads to consistent description of the phases of the model \cite{Jackiw3}. 
The absence of SSB in the large-$N$ ground state was also
argued in \cite{Abbott}:
using RG invariant quantities, a stable symmetric saddle point was 
supported in 4-d, 
albeit with imaginary contributions to the effective potential.
Further studies \cite{halpern}, \cite{brezin}, 
of the large-$N$ saddle point in the $\epsilon$-expansion 
concluded that a second order 
phase transition is present in $d > 2$ dimensions. In addition, it was shown 
that the $O(N)$ invariant mass vanishes at the critical point with a 
critical exponent $\nu = 1/(d-2)$, for $2 < d< 4$ dimensions.

Having in mind the convexity as the principal characteristic of the 
effective potential to hold in a SSB theory, we focus 
in the present article to a complex scalar field 
as well as the $O(N)$-symmetric model, 
and show that the Maxwell construction arises 
from the summation over all the saddle points which constitute a valley of 
minima, and dominate the partition function. This 
semi-classical approximation leads to a convex effective potential, which for large volumes is universal, in the sense that it 
does not depend on the coupling constant of the bare theory.
The potential becomes flat in the limit of infinite volume.
We believe that, for a SSB potential, it is essential to first calculate the partition function, 
taking into account the whole set of minima, and then derive a convex effective action via the Legendre transform.

We describe in section 2 the detailed steps of the quantization of a complex scalar model which features SSB. 
We first show convexity as a general property, and we define our semi-classical approximation for the calculation
of the partition function. We then derive the effective potential and show how the
Maxwell construction occurs in the region where the scalar field modulus $\rho$ is 
smaller than the vev $v$. The semi-classical approximation applied to the outside region $\rho>v$ leads to an effective 
potential identical to the bare potential, which is convex in this region.
Section 3 generalizes the resulting Maxwell construction for an $O(N)$ model, for which the steps are similar 
to those followed for the complex scalar field.
Note that the Maxwell construction obtained here is a result of the full quantization of the $N$ degrees of freedom, unlike 
the effective field theory approach for a linear sigma model,
which consists in integrating the massive degree of freedom only, in order to describe the infrared dynamics of the
remaining Goldstone modes. The latter procedure leads to a non-trivial effective theory, as nicely reviewed in \cite{donoghue}.
In the Appendix, we treat the real field potential with two equivalent vacua $(Z2)$ 
in the same approximation, and we demonstrate that the form of the effective action is reproduced
from the $O(N)$ action when $N=1$.
Finally, section 4 comments on the Abelian Higgs model, where no Goldstone mode is present because of gauge fixing, 
such that the partition function is dominated by one saddle point only. The semi-classical approximation is then 
equivalent to a tree-level approximation, and the usual Higgs mechanism occurs. We also explain why the general argument
of convexity does not hold in the presence of a vector field.

\section{Self-interacting complex scalar field}

\subsection{Construction of the effective action}

We review here the basic features of path integral quantization, in order to introduce our notations.\\
We consider a model with a self-interacting complex scalar field $\phi=\rho\exp(i\alpha)$, 
in which the bare potential $U_{bare}$ depends on $\rho=\sqrt{\phi\phi^\star}$ only.
The partition function is, using a Euclidean metric,
\be\label{Z}
Z[j,j^\star]=\int{\cal D}[\phi,\phi^\star]\exp\left(-S[\phi,\phi^\star]-\int_x j\phi+j^\star\phi^\star \right)~, 
\ee
where $j=re^{i\theta},j^\star=re^{-i\theta}$ are source which parametrizes the system, and which will eventually be 
replaced by the classical fields $\phi_c,\phi_c^\star$, 
defined as
\bea\label{phic}
\phi_c&\equiv&\rho_c e^{i\alpha_c}=-\frac{1}{Z}\frac{\delta Z}{\delta j}=-\frac{e^{-i\theta}}{2Z}\frac{\delta Z}{\delta r}\\
\phi_c^\star&\equiv&\rho_c e^{-i\alpha_c}=-\frac{1}{Z}\frac{\delta Z}{\delta j^\star}
=-\frac{e^{i\theta}}{2Z}\frac{\delta Z}{\delta r}\nonumber~.
\eea
In terms of the polar coordinates $(\rho_c, \alpha_c)$, the above definitions are equivalent to
\be\label{phicpol}
\rho_c = \frac{1}{2Z}\left|\frac{\delta Z}{\delta r}\right| \;\;\;,\;\;\;
\alpha_c =-\theta + \frac{\pi}{2} \left(1
+ {\rm sign}(\delta Z/\delta r)\right)~.  
\ee
We note that the partition function $Z$ depends only on the modulus of the source $j$. Indeed, the source term can be written
\be
\int_x j\phi+j^\star\phi^\star=2\int_x r\rho\cos(\alpha+\theta)~,
\ee
and the summation over all the configurations $\phi$ implies that, for a fixed source $j$, one can change the variable 
$\alpha\to\alpha-\theta$, such that $Z[j,j^\star]=Z[r]$. Nevertheless, we keep explicit the $\theta$ dependence, 
in order to take into account the two degrees of freedom present in the model.\\
The effective action $\Gamma$ is defined as the Legendre transform of $W[r]=-\ln(Z[r])$ with respect to the sources $j,j^\star$
\be\label{Gamma}
\Gamma[\phi_c,\phi_c^\star]=W[r]-\int_x j\phi_c+j^\star\phi_c^\star~,
\ee
where the sources have to be understood as a functionals of the classical fields, after inverting the relations (\ref{phic}).
From the definition (\ref{Gamma}), one finds that the equations of motion for the classical fields are
\be\label{equamotphic}
\frac{\delta\Gamma}{\delta\phi_c}=-j~ \;\;,\;\;
\frac{\delta\Gamma}{\delta\phi_c^\star}=-j^\star ~. 
\ee
We are interested in the effective potential of the theory, which is  
obtained from the momentum independent part of the effective action 
\be
U_{eff}(\rho_c)=\frac{1}{V}\Gamma[\phi_c,\phi_c^\star] ~,~~\mbox{with}~\phi_c=~\mbox{constant}~,
\ee
where $V$ is the volume of space time. In this case, the effective action 
depends on the modulus $\rho_c$ only, and,
taking into account the equations of motion (\ref{equamotphic}), we obtain
\be\label{equamotphicbis}
\frac{1}{2}\left|\frac{\delta\Gamma}{\delta\rho_c}\right|=r~.
\ee
Finally, for constant fields, the functional derivatives become
\be
\frac{\delta(\cdots)}{\delta r}\to\frac{1}{V}\frac{\partial(\cdots)}{\partial r}~~\mbox{and}~~
\frac{\delta(\cdots)}{\delta\rho_c}\to\frac{1}{V}\frac{\partial(\cdots)}{\partial\rho_c}~.
\ee

\subsection{Convexity of the effective potential}\label{convex}

The convexity of the effective action $\Gamma$ is a consequence of its definition as the Legendre transform of the connected graph generating
functional $W[j,j^\star]$, as we explain here.\\
Let us define the operator
\be
\delta^2 W\equiv\left(
\begin{array}{cc}\frac{\delta^2 W}{\delta j \delta j^\star}&
\frac{\delta^2 W}{\delta j \delta j}\\
\frac{\delta^2 W}{\delta j^\star \delta j^\star}
&\frac{\delta^2 W}{\delta j^\star \delta j}\end{array}
\right) ~,
\label{d2W}
\ee 
with the functional derivatives applied at a pair of spacetime points $x,y$
as e.g.
\be
\frac{\delta^2W}{\delta j(x)\delta j^\star(y)}
=\phi_c(x)\phi_c^\star(y)-\left<\phi (x)\phi^\star(y)\right>~,
\label{d2W1}
\ee
where
\be
\left<(\cdots)\right>\equiv\frac{1}{Z}\int{\cal D}[\phi,\phi^\star](\cdots)\exp\left(-S[\phi,\phi^\star]-\int_x j\phi+j^\star\phi^\star \right)~.
\ee
Using the invariance of the Euclidean action under translations and $O(4)$ rotations, the distribution~(\ref{d2W1}) is a real function of $|x-y|$. In addition, it 
is also the opposite of a variance and therefore the diagonal elements of 
the Hermitian operator (\ref{d2W}) are equal and negative. The eigenvalues of
this operator are 
\be
\frac{\delta^2W}{\delta j\delta j^\star}\pm
\left|\frac{\delta^2W}{\delta j\delta j}\right|~,
\label{eig}
\ee
and are negative, since they can be written in terms of variances as
\be
-{\rm var}({\rm Re}\{\phi\}) - {\rm var}({\rm Im}\{\phi\}) \pm
\sqrt{ {\rm var}^2({\rm Re}\{\phi\}) + {\rm var}^2({\rm Im}\{\phi\})}~.
\ee
As a consequence, $W$ is a {\it concave} functional.\\
In order to study the properties of the effective action $\Gamma$, we introduce
the operator 
\be
\delta^2\Gamma\equiv\left(
\begin{array}{cc}\frac{\delta^2\Gamma}{\delta\phi_c^\star \delta\phi_c}&
\frac{\delta^2\Gamma}{\delta\phi_c^\star\delta\phi_c^\star}\\
\frac{\delta^2\Gamma}{\delta\phi_c\delta\phi_c}&\frac{\delta^2\Gamma}{\delta\phi_c\delta\phi_c^\star}\end{array}
\right) ~.
\ee 
Taking into account the definitions~(\ref{phic}), and the equations of 
motion~(\ref{equamotphic}), the operators $\delta^2W$ and $\delta^2 \Gamma$
are:
\be
\delta^2 W = \left(
\begin{array}{cc}\frac{\delta \phi_c^\star}{\delta j}&
\frac{\delta \phi_c}{\delta j }\\
\frac{\delta \phi_c^\star}{\delta j^\star}
&\frac{\delta \phi_c}{\delta j^\star}\end{array}
\right) \;\;\;\; , \;\;\;\;
\delta^2 \Gamma = -\left(
\begin{array}{cc}\frac{\delta j}{\delta \phi_c^\star}&
\frac{\delta j^\star}{\delta \phi_c^\star}\\
\frac{\delta j }{\delta \phi_c}
&\frac{\delta j^\star}{\delta \phi_c}\end{array}
\right) ~,
\ee 
and it can easily be seen that they satisfy the relation\footnote{We remind that $j$ and $j^\star$ are independent variables. Hence diagonal elements
of the product $\delta^2W\cdot\delta^2\Gamma$ involve $\int(\delta j/\delta\phi)(\delta \phi/\delta j^\star)=\delta j/\delta j^\star=0$. 
Similarly, we also have $\int(\delta\phi/\delta j)(\delta j/\delta\phi^\star)=\delta\phi/\delta\phi^\star=0$.}
\be
\delta^2 W \cdot \delta^2 \Gamma = -2 \times {\bf 1}{\bf {\hspace{-0.16cm}1}}~,
\ee
where ${\bf 1}{\bf {\hspace{-0.16cm}1}}$ is the unit operator. $\delta^2\Gamma$ is therefore 
proportional to the inverse of $\delta^2W$, and has positive eigenvalues: $\Gamma$ is a {\it convex} functional of the classical field.\\ 
To see the consequence for the effective potential $U_{eff}(\rho_c)$, a constant configuration for the scalar field 
is enough, and we have
\bea\label{eigen}
\frac{\delta^2}{\delta\phi_c\delta\phi_c^\star}\int d^4x~U_{eff}(\rho_c)&=&\frac{1}{4}\left( U_{eff}''
+\frac{1}{\rho_c}U_{eff}'\right) \delta^4(x-y)\\
\frac{\delta^2}{\delta\phi_c\delta\phi_c}\int d^4x~U_{eff}(\rho_c)&=&\frac{(\phi^\star)^2}{4\rho_c^2}
\left( U_{eff}''-\frac{1}{\rho_c}U_{eff}'\right)\delta^4(x-y)\nn
\frac{\delta^2}{\delta\phi_c^\star\delta\phi_c^\star}\int d^4x~U_{eff}(\rho_c)&=&\frac{\phi^2}{4\rho_c^2}
\left( U_{eff}''-\frac{1}{\rho_c}U_{eff}'\right)\delta^4(x-y)\nonumber~,
\eea 
where a prime denotes a derivative with respect to $\rho_c$. It is then straightforward to calculate
the eigenvalues of $\delta^2\Gamma$, which are $U_{eff}''/2$ 
and $U_{eff}'/(2\rho_c)$. 
The convexity of $\Gamma$ implies then that the effective potential
is necessarily an increasing ($U_{eff}'\ge0$) and convex ($U_{eff}''\ge0$) 
function of $\rho_c$.

\subsection{Semi-classical approximation}

\begin{figure}[ht]
\vspace{-1.7cm}
\begin{center}
\includegraphics[width=12cm]{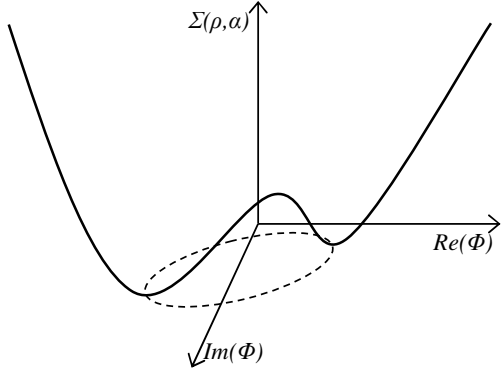}
\caption{The function $\Sigma(\rho,\alpha)$, given in eq.(\ref{sigma1}),  
for external source module $r < r_{crit}$. 
The 'tilted Mexican hat' shape possesses a valley of minima which define the semiclassical 
approximation.}
\label{saddle1}
\end{center}
\end{figure}

\begin{figure}[ht]
\vspace{-2cm}
\begin{center}
\includegraphics[width=12cm]{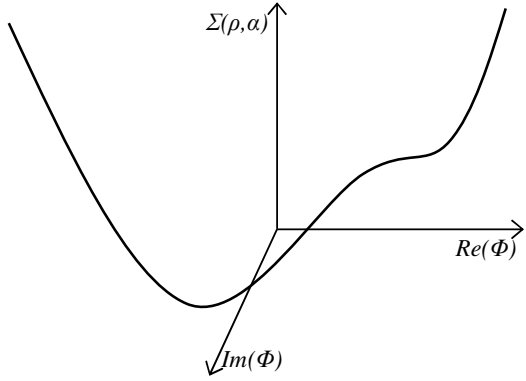}
\caption{The function $\Sigma(\rho,\alpha)$, given in eq.(\ref{sigma1}),
for external source module $r > r_{crit}$. The minima are located sharply around the 
$\cos(\alpha+\theta)=-1$ point. The second local minimum located at 
$\cos(\alpha+\theta)=1$ degenerates to an inflection point at $r = r_{crit}$ and
disappears for $r > r_{crit}$. }
\label{saddle2}
\end{center}
\end{figure}

From now on we consider the following symmetry breaking potential  
\be
U_{bare}(\rho)=\frac{\lambda}{24}(\rho^2-v^2)^2~,
\label{Ubare}
\ee
with a minimum at $\rho = v$.
The semi-classical approximation for the partition function (\ref{Z}) consists in taking its dominant contribution only,
arising from the minima of the functional
\be\label{sigma}
\Sigma[\phi,\phi^*]=S[\phi,\phi^\star]+\int_x j\phi+j^\star\phi^\star \;.
\ee
Since we are interested in the effective potential, we consider only 
homogeneous sources, for which the minima of $\Sigma$ are homogeneous 
fields. Non-homogeneous fields of solitonic type can also contribute to the
partition function but since their action is finite, their contribution is 
negligible compared to the one of homogeneous configurations.
We are therefore interested in the minima of 
\be\label{sigma1}
\Sigma(\rho,\alpha)\equiv
V\left[U_{bare}(\rho)+2r\rho\cos(\alpha+\theta) \right] ~,
\ee
for a given angle $\alpha+\theta$ and modulus $r$. One therefore looks for the real 
and positive solution $\rho_0$ of the equation
\be
\frac{\lambda}{6}(\rho_0^2-v^2)\rho_0+2r\cos(\alpha+\theta)=0~,
\label{cubic}
\ee
which goes to $v$ when the source term $r\cos(\alpha+\theta)$ vanishes. 
Defining the critical source modulus 
\be\label{rcrit}
r_{crit}=\frac{\lambda v^3}{18\sqrt{3}}~,
\ee
we will study independently the two cases:
\begin{itemize}
\item $r \le r_{crit}$: Equation~(\ref{cubic}) has three real solutions, 
among which the local minimum is given by
\be\label{rho0}
\rho_0=\frac{2v}{\sqrt 3}\cos\left\lbrace\frac{\pi}{3}-\frac{1}{3}
\arccos\left(\frac{r}{r_{crit}}\cos (\alpha+\theta)\right)\right\rbrace ~.
\ee
The function $\Sigma(\rho,\alpha)$ takes the form of a 'tilted Mexican hat 
potential' (Fig.~\ref{saddle1}),
with the set of corresponding saddle points forming a valley.
The absolute minimum of the resulting valley corresponds to
$\cos(\alpha+\theta)=-1$, while the top of the valley is located at $\cos(\alpha+\theta)=1$. 
It is also easy to verify that the local maximum $\rho_1$ of $\Sigma(\rho,\alpha)$, which goes to 0 when the source vanishes, 
reads
\be
\rho_1=\frac{2v}{\sqrt 3}\cos\left\lbrace\frac{\pi}{3}+\frac{1}{3}
\arccos\left(\frac{r}{r_{crit}}\cos (\alpha+\theta)\right)\right\rbrace ~.
\ee
An equivalent criterion for the existence of the valley is that its 
top is further away than the local maximum in the radial direction, i.e. 
$\rho_1<\rho_0$ for $\cos(\alpha+\theta)=1$. 
This condition is indeed satisfied if $r<r_{crit}$. Notice that when 
$r = r_{crit}$, the top of the valley point merges with the local maximum
$\rho_1$ and becomes an inflection point (Fig.~\ref{saddle2}).

\item $r > r_{crit}$: The equation~(\ref{cubic}) has the real and positive 
solution~(\ref{rho0}) for restricted values of $\alpha$, which satisfy   
$| \cos(\alpha+\theta)| \le r_{crit}/r$ (see Fig.~\ref{saddle2}). 
Nevertheless, this set of local minima are much higher than the absolute 
minimum which is obtained for $\cos(\alpha+\theta)=-1$ and which reads  
\be\label{rho0general}
\rho_0=\frac{2v}{\sqrt 3}\cosh\left\lbrace\frac{1}{3}
\cosh^{-1}\left( \frac{r}{r_{crit}}\right)\right\rbrace ~.
\ee
Since $\Sigma$ sharply deepens in the neighborhood of the absolute minimum (\ref{rho0general}), 
the semi-classical approximation consists in taking into account the contribution of the latter point  only.

\end{itemize}

\subsection{Maxwell construction}\label{Maxwell}

As discussed in the previous subsection, in the case where $r<r_{crit}$ the partition function is dominated by:
\bea
Z[r]&\simeq&\int_0^{2\pi}\frac{d\alpha}{2\pi} \exp\left(-V\left[U_{bare}(\rho_0)+2r\rho_0\cos(\alpha+\theta) \right]\right)\\
&=&\int_0^{2\pi}\frac{d\alpha}{2\pi} \exp\left(-V\left[U_{bare}(\rho_0)+2r\rho_0\cos\alpha \right]\right)~,\nonumber
\eea
where $\rho_0$ depends on the source $r$ and is given by eq.(\ref{rho0}).
We note that the summation over all the points in the valley ensures that the 
partition function $Z$ depends on the modulus $r$ of the source only.
After introducing the dimensionless quantities
\be
A\equiv\lambda\frac{Vv^4}{24}~~,~~\tilde\rho_0\equiv\frac{\rho_0}{v}~~,~~\tilde r\equiv\frac{r}{r_{crit}}~,
\ee
the minimum $\tilde\rho_0$ can be expanded in powers of $\tilde r$, and we find, up to fourth order,
\be
\tilde\rho_0=1-\frac{\sqrt3}{9}\tilde r\cos\alpha-\frac{1}{18}(\tilde r\cos\alpha)^2-\frac{4\sqrt3}{243}(\tilde r\cos\alpha)^3
-\frac{35}{1944}(\tilde r\cos\alpha)^4+{\cal O}(\tilde r^5)~.
\ee
We obtain then
\bea
&&\exp\left(-V\left[U_{bare}(\rho_0)+2r\rho_0\cos\alpha \right]\right)\\
&=&1-\frac{8\sqrt3}{9}A\tilde r\cos\alpha+\frac{4}{27}A\left( 1+8A\right) (\tilde r\cos\alpha)^2\nn
&+&\frac{4\sqrt3}{243}A\left(1-8A-\frac{64}{3}A^2 \right) (\tilde r\cos\alpha)^3\nn
&+&\frac{8}{243}A\left(\frac{1}{3}-A+\frac{16}{3}A^2+\frac{64}{9}A^3 \right) (\tilde r\cos\alpha)^4+{\cal O}(\tilde r^6)~,
\eea
and the integration over $\alpha$ leads finally to
\be\label{Ztilder}
Z(\tilde r)=1+\frac{2A}{27}\left(1+8A\right) \tilde r^2+\frac{A}{81}\left(\frac{1}{3}-A+\frac{16}{3}A^2+\frac{64}{9}A^3 \right)
\tilde r^4+{\cal O}(\tilde r^6)~.
\ee
This last expression will now be used to calculate the effective potential, around the origin.\\
From the definition (\ref{phicpol}) of the classical field, one finds
\be
\rho_c=v\frac{3\sqrt3}{8AZ}\frac{\partial Z}{\partial\tilde r}~,~~~~~\alpha_c=\pi-\theta~, 
\ee 
such that the expansion (\ref{Ztilder}) leads to
\be
\tilde\rho_c\equiv\frac{\rho_c}{v}=\frac{\sqrt3}{18}\left(1+8A\right)\tilde r
+\frac{\sqrt3}{81}\left( \frac{1}{2}-\frac{11}{6}A+\frac{8}{3}A^2-\frac{32}{3}A^3\right) \tilde r^3 +{\cal O}(\tilde r^5)~.
\ee
One can see that the regime $r<r_{crit}$ leads to a polynomial expansion of the classical field around $\rho_c=0$.
We then invert the last series, by expanding $\tilde r$ in powers of $\tilde\rho_c$, and we find
\be\label{tilder}
\tilde r=\frac{18}{\sqrt3(1+8A)}~\tilde\rho_c+\frac{144\sqrt3(-1/2+11A/6-8A^2/3+32A^3/3)}{(1+8A)^4}~\tilde\rho_c^3+{\cal O}(\tilde\rho^5)~.
\ee
Since $\Gamma=VU_{eff}$ is an increasing function of $\rho_c$, the equation of motion (\ref{equamotphicbis}) can also be written
\be
\left|\frac{\partial\Gamma}{\partial\tilde\rho_c}\right|=\frac{\partial\Gamma}{\partial\tilde\rho_c}=\frac{8A}{3\sqrt3}~\tilde r~,
\ee
Together with the expansion (\ref{tilder}), the integration over $\tilde\rho_c$ gives
\be\label{convexG}
\Gamma[\tilde\rho_c]=\frac{8A}{1+8A}\tilde\rho_c^2+\frac{48A(-1+11A/3-16A^2/3+64A^3/3)}{(1+8A)^4}\tilde\rho_c^4+{\cal O}(\tilde\rho_c^6)~,
\ee
where the constant of integration is disregarded.
In the limit of large volume $A>>1$, the effective action is then
\be
\Gamma[\rho_c]=\left(\frac{\rho_c}{v} \right)^2+\frac{1}{4} \left(\frac{\rho_c}{v} \right)^4+\cdots~,
\ee
and the effective potential is finally obtained after dividing by the volume
\be\label{Ueffinal}
U_{eff}(\rho_c)=\frac{1}{V}\Gamma[\rho_c]=
\frac{1}{V}\left( \frac{\rho_c}{v}\right)^2+\frac{1}{4V} \left(\frac{\rho_c}{v} \right)^4 +\cdots~.
\ee
$U_{eff}$ therefore vanishes in the limit of infinite volume
\be
U_{eff}(\rho_c)\to0~~~~\mbox{(infinite volume)}~,
\ee
and has the form of a flat disc for $\rho_c<<v$. Few interesting remarks can be made here:

\begin{itemize}
\item This study has been done up to the 4th order in the classical field, but
it is clear that the present construction can be extended to any order in $\rho_c$. 
Together with the convexity of the effective potential, proven in subsection \ref{convex}, we expect 
the flatness to hold up to $\rho_c\simeq v$.

\item The convexity of the potential (\ref{Ueffinal}) is actually valid for finite volume, and not only in the limit of infinite volume;

\item The effective potential (\ref{Ueffinal}) is universal and does not depend on the
bare self-coupling of the scalar field. This universality has been observed in exact Wilsonian renormalization group studies, where
the Maxwell construction is obtained in the infrared limit of the so-called spinodal region \cite{spinodal}. 
In this region, the running potential is a universal
quadratic function of the background field, independent of the details of the spontaneously broken ultraviolet potential. 
\end{itemize}
To conclude this subsection, quantization of the theory
erases the non-convex part of the bare potential, as expected from the general argument given in subsection \ref{convex},
and verified within our semi-classical approximation.

\subsection{Semi-classical partition function for $\rho_c>v$}\label{rho>v}

In the situation where $r>r_{crit}$, 
there is no valley of saddle points anymore, but only one saddle point, which is obtained for $\cos(\alpha+\theta)=-1$, 
and it is easy to show that the semi-classical approximation leads to an effective potential which is identical 
to the bare potential, as we do here. The approximate partition function is
\be
Z\simeq\exp\left(-V(U_{bare}(\rho_0)-2r\rho_0) \right)~,
\ee
where $\rho_0$ is given by eq.(\ref{rho0general}). The classical field is therefore
\be
\phi_c=\frac{-e^{-i\theta}}{2V}\frac{\partial(\ln Z)}{\partial r}
=\frac{e^{-i\theta}}{2}\left[ \left( \frac{dU_{bare}}{d\rho_0}-2r\right)\frac{d\rho_0}{dr}-2\rho_0\right] = e^{i(\pi-\theta)}\rho_0(r)~,
\ee
such that 
\be\label{phicbis}
\rho_c=\rho_0(r)~~~~\mbox{and}~~~~\alpha_c=\pi-\theta~.
\ee
In the semi-classical approximation, the partition function can then be expressed as
\be
Z=\exp\left(-V(U_{bare}(\rho_c)-2r\rho_c )\right)~, 
\ee
and, according to the result (\ref{phicbis}), the effective potential is finally
\be
U_{eff}(\rho_c)=-\frac{1}{V}\left( \ln Z+2Vr\rho_c\cos(\alpha_c+\theta)\right) =U_{bare}(\rho_c)~.
\ee
Note that this result would also be obtained for a bare potential which does not feature spontaneous symmetry breaking. In that sense, 
the saddle point approximation (although non-perturbative in principle)
constitutes a 'tree level' approximation for potentials with a unique vacuum,
or, far away from the degenerate vacua region. On the other hand, 
since it is sensitive to the presence of degenerate vacua, the saddle point
summation captures a genuine non-perturbative effect which is the flattening of
the potential within the disc of $\rho_c < v$. The smooth matching of the
potential at $\rho_c \sim v$ between the flat disc and the 
convex outer branch would of course require a full non-perturbative 
calculation, currently possible only via  Lattice Field Theory techniques.

\section{Effective Potential for $O(N)$-symmetric theories}

\subsection{Convexity of the Effective Potential}

It is straightforward to generalize the results of the previous section to
scalar theories with a global $O(N)$ symmetry. For the N-plet of fields
$\vec{\phi}=(\phi_1,\phi_2,\dots,\phi_N)$ the action reads in Euclidean space
\be
S[\vp]=\int d^4x \left\{ \frac{1}{2}\partial_\mu \vp \cdot \partial_\mu \vp+ 
\frac{\lambda}{4}\left(\rho^2-v^2\right)^2 \right\} ~,
\ee
where $\rho=\sqrt{\vp\cdot\vp}$. 
The partition function is
\be
Z[\vec{j}] = \int {\cal D}[\vp] ~ \exp \left(-S[\vp] -\int_x \vec{j}\cdot \vp 
\right)~,
\label{ZN}
\ee
where $\vec{j}$ is the $O(N)$ vector source coupled to the fields.
$Z$ and the connected graphs generating functional $W= -\ln Z$
depend only on the modulus $r=|\vec{j}|$ of the source vector as a 
consequence of the $O(N)$ invariance of the action and the 
integration measure ${\cal D}\vp$. 
The classical field $\vec{\phi_c}$ is defined via 
\be
\vpc = -\frac{1}{Z}\frac{\delta Z}{\delta \vec{j}} = 
\frac{\delta W}{\delta \vec{j}} = \frac{\vec{j}}{r}~\frac{\delta W}{\delta r}  ~,
\label{phiN}
\ee
and therefore its modulus $\rho_c = |\vpc|$ is
\be\label{phiNmod}
\rho_c = \frac{1}{Z}\left|\frac{\delta Z}{\delta r}\right|~.
\ee
The effective action is defined via the Legendre transform
\be
\Gamma[\vpc]=W[r]-\int_x \vec{j}\cdot \vpc ~,
\ee
such that 
\be
\frac{\delta \Gamma}{\delta \vpc} = -\vec{j} ~~~~~{\rm and} ~~~~~
\left|\frac{\delta \Gamma}{\delta \rho_c}\right| = r ~.  
\label{eomN}
\ee
The generating functional $W$ is concave, which can be shown as follows.  
We define the $N\times N$ operator with matrix elements
\be
\left(\delta^2 W\right)^{ab} \equiv \frac{\delta ^2 W}{\delta j_a \delta j_b} = 
\frac{\delta \phi_c^a}{\delta j_b }=
\phi_c^a \phi_c^b -\left< \phi^a \phi^b\right>
\ee
where $\left<\dots \right>$ denotes an expectation value with respect to the
partition function~(\ref{ZN}).
Since $-\delta^2 W$ coincides with the covariance matrix for 
the N-plet of fields $\vp$, it has the general property of being
a positive semi-definite matrix, with eigenvalues greater than zero.
Following the steps in section 2, the convexity of $\Gamma$ arises from the
examination of the operator $\delta^2 \Gamma$ with matrix elements
\be
\left(\delta^2 \Gamma \right)_{ab} \equiv \frac{\delta ^2 \Gamma}
{\delta \phi_c^a \delta \phi_c^b} = -\frac{\delta j_a}{\delta \phi_c^b}~,
\ee
since
\be
\delta^2 W \cdot \delta^2 \Gamma = -{\bf 1}{\bf {\hspace{-0.16cm}1}}~,
\ee
where ${\bf 1}{\bf {\hspace{-0.16cm}1}}$ denotes the $N \times N$ unit matrix.
For the effective potential $U_{eff}(\rho_c)$ we consider constant field 
configurations such that
\be
\frac{\delta ^2}{\delta \phi_c^a \delta \phi_c^b} \int d^4x~ U_{eff}(\rho_c) =
\left[ \frac{U'_{eff}}{\rho_c}~ \delta^{ab} + 
 \left(U_{eff}'' - \frac{U'_{eff}}{\rho_c} \right)
\frac{\phi_c^a \phi_c^b}{\rho_c^2}
\right]\delta^4(x-y)~.
\ee
Algebraically it is straightforward to check that an $N\times N$ real matrix
of the form 
\be
M^{ab} = A~ \delta^{ab} + B~ n^a n^b ~~~,~{\rm with}~~~ \sum_{a=1}^N n^a n^a =1~,
\ee
possesses the eigenvalues\footnote{Consider for example an $O(N)$ rotation of 
the unit vector $\vec{n}\rightarrow (0,0,...,0,1)$ which brings $M$ 
 to the form $diag (A,A,...,A,A+B)$.}
\be
\lambda_1=\lambda_2=\dots=\lambda_{N-1} = A  ~~~{\rm and}~~\lambda_N = A+B~.
\ee
The convexity of $\Gamma$ therefore guarantees that
\be
U'_{eff}(\rho_c) \ge 0 ~~~{\rm and}~~~  U''_{eff}(\rho_c) \ge 0~,
\ee
and the effective potential is an increasing and convex function of $\rho_c$.

\subsection{Maxwell Construction}

The semiclassical approximation to the partition function~(\ref{ZN}) consists 
of summing over the minima of the functional
\be
\Sigma[\vec{\phi}]=S[\vec{\phi}] + \int_x \vec{j}\cdot \vec{\phi}~.
\ee
For the effective potential it suffices to consider constant field 
configurations and minimize the function 
\be
\Sigma[\rho,\omega] = V \left[\frac{\lambda}{4}\left(\rho^2-v^2\right)^2 +
r\rho  \cos \omega \right]~,
\ee
where $\omega$ is the angle $(0\le \omega \le \pi)$ between $\vec{j}$ and
$\vp$ in field space. Thus, we search for the real and positive solutions 
$\rho_0$ of
\be
\lambda (\rho^2_0 - v^2) \rho_0 = -r \cos\omega ~.  
\ee
and the analysis presented in subsection (2.3) holds for a critical source
modulus now taking the value
\be
r_{crit}=\frac{2\lambda v^3}{3\sqrt{3}}~.
\ee
For $r \le r_{crit}$, $\Sigma$ takes the shape of a 
'tilted Mexican-hat potential' with the minima given by
\be
\rho_0=\frac{2v}{\sqrt 3}\cos\left\lbrace\frac{\pi}{3}-\frac{1}{3}
\arccos\left(\frac{r}{r_{crit}}\cos \omega\right)\right\rbrace ~.
\ee
and the saddle point approximation consists of integrating over the 
$N-$ dimensional solid angle $\Omega_{N-1}$,
which covers the valley of radial minima
$\vp^2 = \rho_0^2$ in field space: 
\be
Z[r]\simeq  \int \frac{d\Omega_{N-1}}{\Omega_{N-1}}\exp\left(-\Sigma[\rho_0,\omega] \right)~.
\ee
Rescaling variables as 
\be
A\equiv\lambda\frac{Vv^4}{4}~~,~~\tilde\rho_0\equiv\frac{\rho_0}{v}~~,~~\tilde r\equiv\frac{r}{r_{crit}}~,
\ee
we obtain, as in the complex scalar case,
\bea\label{expSigmaN}
\exp(-\Sigma[\rho_0,\omega])&=&1-\frac{8\sqrt3}{9}A\tilde r\cos\omega+\frac{4}{27}A\left( 1+8A\right) (\tilde r\cos\omega)^2\\
&+&\frac{4\sqrt3}{243}A\left(1-8A-\frac{64}{3}A^2 \right) (\tilde r\cos\omega)^3\nn
&+&\frac{8}{243}A\left(\frac{1}{3}-A+\frac{16}{3}A^2+\frac{64}{9}A^3 \right) (\tilde r\cos\omega)^4+{\cal O}(\tilde r^6)~.\nonumber
\eea
Using the $O(N)$ invariance of $Z$ in the internal space, we can rotate
$\vec{j}$ on the $N-$th axis such that 
$\omega$ becomes the polar angle of $\vp$
with the $N-$th axis, thus effectively reducing
\be
\frac{d\Omega_{N-1}}{\Omega_{N-1}}~ \longrightarrow~
\frac{d\omega}{I_{N-2}}(\sin\omega)^{N-2}~~,~\mbox{with}~0\leq\omega\leq\pi~,
\ee
where we define
\be
I_n=\int_0^\pi d\omega(\sin\omega)^n~.
\ee
These integrals satisfy
\be
I_n=\frac{n-1}{n}I_{n-2}~,
\ee
and the calculation of the partition function involves only the even powers 
of $\cos\omega$ in the expression (\ref{expSigmaN}):
\bea
\int d\omega(\sin\omega)^{N-2}&=&I_{N-2}\\
\int d\omega(\sin\omega)^{N-2}\cos^2\omega&=&I_{N-2}-I_N=\frac{I_{N-2}}{N}\nn
\int d\omega(\sin\omega)^{N-2}\cos^4\omega&=&I_{N-2}-2I_N+I_{N+2}=\frac{3I_{N-2}}{N(N+2)}~.\nonumber
\eea
As a consequence, the partition function is
\be
Z(\tilde r)=1+\frac{4A}{27N}\left(1+8A\right) \tilde r^2
+\frac{8A}{81N(N+2)}\left(\frac{1}{3}-A+\frac{16}{3}A^2+\frac{64}{9}A^3 \right)
\tilde r^4+{\cal O}(\tilde r^6)~.
\ee
From the classical modulus (\ref{phiNmod}), we find then
\bea
\tilde\rho_c&\equiv&\frac{\rho_c}{v}=\frac{3\sqrt3}{8AZ}\frac{\partial Z}{\partial\tilde r}
=\frac{\sqrt3}{9N}\left(1+8A\right)\tilde r+\\
&&\frac{-4\sqrt3}{243N^2(N+2)}\left(-3N+2(1+5N)A+32(1-N)A^2+128A^3\right) \tilde r^3 +{\cal O}(\tilde r^5)~,\nonumber
\eea
and the corresponding expansion of $\tilde{r}$ in terms of $\tilde\rho_c$ reads
\bea
\label{rN}
\tilde r&=&\frac{9N}{\sqrt3(1+8A)}~\tilde\rho_c+\\
&&\frac{12\sqrt3N^2}{(N+2)(1+8A)^4}(-3N+2(1+5N)A+32(1-N)A^2+128A^3)
~\tilde\rho_c^3+{\cal O}(\tilde\rho^5)~.\nonumber
\eea
Since $\Gamma=VU_{eff}$ is an increasing function of $\rho_c$, the equation of motion (\ref{eomN}) can also be written
\be
\left|\frac{\partial\Gamma}{\partial\tilde\rho_c}\right|=\frac{\partial\Gamma}{\partial\tilde\rho_c}=\frac{8A}{3\sqrt3}~\tilde r~,
\ee
Together with the expansion (\ref{rN}), the integration over $\tilde\rho_c$ gives
\bea
\label{ONgamma}
\Gamma[\tilde\rho_c]&=&\frac{4NA}{1+8A}\tilde\rho_c^2\\
&&+\frac{8N^2A}{(N+2)(1+8A)^4}(-3N+2(1+5N)A+32(1-N)A^2+128A^3)\tilde\rho_c^4+{\cal O}(\tilde\rho_c^6)~,\nonumber
\eea
where the constant of integration is disregarded. Note that this result reproduces the effective action (\ref{convexG}) when $N=2$.
In the limit of large volume $A>>1$, the effective action is then
\be
\Gamma[\rho_c]=\frac{N}{2}\left(\frac{\rho_c}{v} \right)^2+\frac{N^2}{4(N+2)} \left(\frac{\rho_c}{v} \right)^4+\cdots~,
\ee
and the effective potential is finally obtained after dividing by the volume
\be
U_{eff}(\rho_c)=\frac{1}{V}\Gamma[\rho_c]=
\frac{N}{2V}\left( \frac{\rho_c}{v}\right)^2+\frac{N^2}{4(N+2)V}\left(\frac{\rho_c}{v} \right)^4 +\cdots~.
\label{UeffN}
\ee
As in the complex scalar case, $U_{eff}$ vanishes in the limit of infinite volume
\be
U_{eff}(\rho_c)\to0~~~~\mbox{(infinite volume)}~,
\ee
and has the form of a flat $N$-ball for $\rho_c<v$. It is interesting to 
speculate about the large-$N$ limit of eq.(\ref{UeffN}). The first two terms of 
the series scale with $N/V$ at large-$N$, indicating a possible symmetry 
restoring vacuum (effective potential convex but not flat, with a minimum at $\rho_c=0$), 
if the large-volume and large-$N$ limit are taken simultaneously, with constant ratio $N/V$. 
In two dimensions, one may argue in favor of the existence of such a limit
as symmetry breaking and massless Goldstone modes are prohibited by
the Coleman-Mermin-Wagner theorem \cite{Coleman}.

Finally, it is clear that for, $r > r_{crit}$, the unique minimum~(\ref{rho0general}) of $\Sigma[\rho,\omega]$
dominates the partition function at the anti-alignment point $\cos\omega =-1$,
and leads to an effective potential identical to the bare one. Such a result
would also hold within the semiclassical approximation for a potential without SSB.

\section{Remarks on the Abelian Higgs model}

In this section, we demonstrate that,
because of gauge fixing, no summation over the phase is involved in the semi-classical approximation for
the partition function of the Abelian Higgs model.
As a consequence, the presence of the gauge field does not allow 
the non-perturbative flattening of
the scalar effective potential.\\
The Abelian Higgs model is described by the Lagrangian (Euclidean metric)
\be
-\frac{1}{4}F_{\mu\nu}F^{\mu\nu}+D_\mu\phi(D^\mu\phi)^\star+U_{bare}(\rho)~,
\ee
where $D_\mu=\partial_\mu-ieA_\mu$ and $U_{bare}$ is the standard SSB 
potential~(\ref{Ubare}).
The covariant derivative term can be written in terms of the modulus $\rho$ and argument $\alpha$ of the scalar field as
\be
D_\mu\phi(D^\mu\phi)^\star=\partial_\mu\rho\partial^\mu\rho+\rho^2\partial_\mu\alpha\partial^\mu\alpha
+e^2\rho^2A_\mu A^\mu-2e\rho^2A^\mu\partial_\mu\alpha~.
\ee
We then perform the gauge transformation 
\be
A_\mu\to A_\mu+e^{-1}\partial_\mu\alpha~~,~~~~\phi\to \rho~,
\ee
which eliminates completely the argument $\alpha$ from the theory
\be
D_\mu\phi(D^\mu\phi)^\star=\partial_\mu\rho\partial^\mu\rho+e^2\rho^2A_\mu A^\mu~.
\ee
As a consequence, 
there is no summation over the scalar field argument, and the partition function is 
dominated by one saddle point only $(\rho=\mbox{const},A_\mu=0)$, leading to the argument similar to the one given in section \ref{rho>v}.
Therefore, based on the semi-classical approximation, the effective potential in the scalar sector remains identical to the bare one.\\
The reason why the convexity argument given in subsection (\ref{convex}) does not hold in this case is precisely gauge fixing, as we now explain.
If one denotes $J^\mu$ the source for the gauge field, the corresponding classical fields of the theory are
\be\label{AmuJmu}
A_\mu^c=\frac{\delta W}{\delta J^\mu}~\;\;\;,\;\;\;
\phi_c = \frac{\delta W}{\delta j} ~\;\;\;,\;\;\;
\phi_c^\star = \frac{\delta W}{\delta j^\star}~\;,
\ee
and the connected graph generating functional $W$ will be a concave functional of the sources $j,j^\star,J^\mu$. 
But in order to define the functional Legendre transform $\Gamma$, one needs to invert simultaneously the relations in (\ref{AmuJmu}) to express 
\be
\Gamma[\phi_c,\phi^\star_c,A_\mu^c]=W[j,j^\star,J^\mu]-\int_x (j\phi_c+j^\star\phi^\star_c+J^\mu A_\mu^c)~,
\ee
where the sources are functionals of the classical fields. 
Since gauge symmetry relates locally the classical fields $A_\mu^c, \phi_c$
and $\phi^\star_c$,
this inversion is possible only after a choice of gauge, which reduces the space of
fields, in order to have a one-to-one relation between sources and a gauge slice of classical fields. As a consequence, the operator 
\be
\delta^2\Gamma=\left(
\begin{array}{ccc}\frac{\delta^2\Gamma}{\delta\phi_c^\star \delta\phi_c}&
\frac{\delta^2\Gamma}{\delta\phi_c^\star\delta\phi_c^\star}&\frac{\delta^2\Gamma}{\delta\phi_c^\star\delta A_\mu^c}\\
\frac{\delta^2\Gamma}{\delta\phi_c\delta\phi_c}&\frac{\delta^2\Gamma}{\delta\phi_c\delta\phi_c^\star}
&\frac{\delta^2\Gamma}{\delta\phi_c\delta A_\mu^c}\\
\frac{\delta^2\Gamma}{\delta\phi_c\delta A_\nu^c}&\frac{\delta^2\Gamma}{\delta\phi_c^\star\delta A_\nu^c}
&\frac{\delta^2\Gamma}{\delta A_\nu^c\delta A_\mu^c}
\end{array}\right) ~,
\ee
acts only 
on a sub-space of fields, compared to the operator 
\be
\delta^2W=\left(
\begin{array}{ccc}\frac{\delta^2 W}{\delta j \delta j^\star}&\frac{\delta^2 W}{\delta j \delta j}&
\frac{\delta^2 W}{\delta j \delta J^\mu}\\
\frac{\delta^2 W}{\delta j^\star \delta j^\star}&\frac{\delta^2 W}{\delta j^\star \delta j}&
\frac{\delta^2 W}{\delta j^\star \delta J^\mu}\\
\frac{\delta^2 W}{\delta j^\star \delta J^\nu}&\frac{\delta^2 W}{\delta j \delta J^\nu}&
\frac{\delta^2 W}{\delta J^\nu \delta J^\mu}
\end{array}\right) ~,
\ee
and cannot be its inverse anymore: the concave properties of $W$ do not lead to the convexity of $\Gamma$. \\
One can consider the following example, which shows the reduction of source space. If the source $\tilde J^\mu$
leads to the classical field $\tilde A_\mu^c$, we have
\be
A_\mu^c=\frac{\delta W}{\delta J^\mu}=\int_x \frac{\delta W}{\delta\tilde J^\nu}\frac{\delta \tilde J^\nu}{\delta J^\mu}
=\int_x \tilde A_\nu^c\frac{\delta \tilde J^\nu}{\delta J^\mu}~.
\ee
By choosing the specific relation
\be
\tilde J^\nu=J^\nu+\theta\partial^\nu\partial_\rho J^\rho~,
\ee
where $\theta$ is any scalar function with mass dimension -2, it is easy to see that $A_\mu^c$ and $\tilde A_\mu^c$ are related by the gauge transformation
\be
A_\mu^c=\tilde A_\mu^c+\partial_\mu\Lambda~~,~\mbox{with}~~\Lambda=\partial^\rho(\theta \tilde A_\rho^c)~.
\ee
Therefore the use of both sources $J^\mu$ and $\tilde J^\mu$ leads to a redundancy of the classical fields, 
and $\tilde J^\mu$ should not be taken into account in the formal inversion of
$\delta^2 W$ which defines $\delta^2 \Gamma$,
if $J^\mu$ is already considered.\\
A complementary argument showing that one cannot prove convexity for the scalar effective potential is the following. One could naively 
think of integrating over the gauge field first, in the path integral of the Abelian Higgs model, in order to obtain an effective theory for the 
scalar field, such that the integration over the scalar field might lead to a convex effective potential. But the integration over 
the gauge field would actually generate a singular effective theory for the scalar field, in such a way that the convexity argument would not hold, because
it is based on functional derivatives of $W$ and $\Gamma$. Indeed, in the above-mentioned gauge where the scalar field is real, the gauge sector
of the bare Abelian Higgs Lagrangian is 
\be
A^\mu{\cal D}^{-1}_{\mu\nu}A^\nu~,~~\mbox{with}~~{\cal D}^{-1}_{\mu\nu}=(\Box+e^2\rho^2)\eta_{\mu\nu}-\partial_\mu\partial_\nu~,
\ee
where the operator ${\cal D}^{-1}_{\mu\nu}$ is invertible for $\rho\ne0$ only.\\
In order to quantize the model, one needs to choose a specific vacuum for the scalar field, and consider radial fluctuations around this vacuum, which 
are represented by a real scalar field. No Maxwell construction arises then, because the partition function is dominated by one minimum only, and the 
Higgs mechanism occurs as expected.

\section{Conclusions}

In this work we generalized the concept of Maxwell construction to scalar field theories
with a continuous group symmetry possessing a non-trivial set of classical 
vacua.
We demonstrated that the effective action is necessarily convex and, within a semiclassical
approximation that takes into account all the degenerate space of  non-trivial vacua, 
quantum fluctuations erase the non-convex part of the action. The result is a flat-disc shaped effective potential, 
for classical field values smaller than the vev of the system.
We stress that it is crucial to first evaluate the partition function, and then perform the Legendre transform in order to 
arrive at this result.
Thus, the vacuum of the quantized theory consists in a superposition 
of states with different field modulus, as a result of the absence of restoration force. 
This mechanism is the analogue of the coexistence of
different phases for a statistical system, during a first order phase transition.\\
We noted that when the complex scalar is coupled to an Abelian gauge field and a Higgs 
mechanism is present, the convexity argument does not hold in general as the 
continuous set of vacua is eliminated by the gauge degrees of freedom. \\
The expression for the convex effective potential found in this work is 
independent of the coupling constant of the bare model, in the large volume limit, up to  
fourth order in the field at least, and depends only on the bare vev.
This universality suggests that the construction presented here is independent of the 
details of the SSB bare potential.\\
A more complete study would involve a loop expansion around the dominant contributions which 
participate in the Maxwell construction.
The flatness of the effective potential would certainly hold, since it 
appears to be the only 
way to satisfy convexity, if one starts with a SSB bare potential.
The loop expansion would help determine the radius of convergence of the expansion 
in terms of the classical field, as well as the transition from the flat regime
to the asymptotic classical form which occurs
at values of the field modulus much larger than the vev.

\appendix

\section*{Appendix: Maxwell construction for a real scalar field with SSB}

The Maxwell construction presented in \cite{alex}, for a real scalar field, is based on a weaker approximation than the one presented here: 
it takes into account the minimum of the bare potential only, not including the source term. The advantage of this approximation is the
possibility to derive an analytical expression for the effective action, without the need for Taylor expansions in the classical field. 
As expected, the corresponding effective potential becomes flat in the limit of infinite volume. The drawback of the approximation in 
\cite{alex} is that the finite-volume expression for the effective action is not accurate, and only the infinite volume limit can be trusted. 
For the sake of completeness, we derive here the Maxwell construction for a real scalar field, using the more appropriate approximation 
described in the present article. We give only the main steps, which are
 similar to those detailed for the complex scalar field. As will be shown, the result coincides with the $O(N)$-symmetric model, with $N=1$, 
 not only for infinite volume, but for any finite volume.\\
The SSB potential in the standard normalization is
\be
U_{bare}(\phi)=\frac{\lambda}{24}(\phi^2-v^2)^2~,
\ee
and the partition function and classical field are 
\be
Z[j]=\int{\cal D}[\phi]\exp\left(-S[\phi]-\int_x j\phi \right)~,
~~\phi_c=-\frac{1}{Z}\frac{\delta Z}{\delta j}\to-\frac{1}{VZ}\frac{\partial Z}{\partial j}~.
\ee
This partition function is dominated by the two saddle points
\be
\phi_+=\frac{2v}{\sqrt 3}\cos\left\lbrace\frac{\pi}{3}-\frac{1}{3}
\arccos\left(\frac{j}{j_{crit}}\right)\right\rbrace ~,
\phi_-=\frac{2v}{\sqrt 3}\cos\left\lbrace\ \pi - \frac{1}{3}
\arccos\left(\frac{j}{j_{crit}}\right)\right\rbrace ~,
\ee
where 
\be
j_{crit}=\frac{\lambda v^3}{9\sqrt3}~,
\ee
and $Z$ can be approximated by
\be
Z[j]\simeq\frac{1}{2}\exp\left(-V[U_{bare}(\phi_+)+j\phi_+]\right)+\frac{1}{2} \exp\left(-V[U_{bare}(\phi_-)+j\phi_-]\right)~.
\ee
An expansion in powers of $j$ gives
\be
Z=1+\frac{4A}{27}(1+8A)\tilde j+\frac{8A}{243}\left( \frac{1}{3}-A+\frac{16}{3}A^2+\frac{64}{9}A^3\right) \tilde j^3+\cdots~,
\ee
where $A=\lambda Vv^4/24$ and dots represent higher orders in $\tilde j=j/j_{crit}$. The classical field is therefore given by
\be
\frac{\phi_c}{v}\equiv\tilde\phi_c=-\frac{\sqrt3}{9}(1+8A)\tilde j-
\frac{4\sqrt3}{729}(3-12A-128A^3)\tilde j^3+\cdots~,
\ee
and the inversion of this expansion gives
\be
\tilde j=-\frac{3\sqrt3}{(1+8A)}\tilde\phi_c-\frac{4\sqrt3}{(1+8A)^4}(-3+12A+128A^3)\tilde\phi_c^3+\cdots~.
\ee
The integration of the equation of motion
\be
\frac{1}{V}\frac{\partial\Gamma}{\partial\phi_c}=-j~,
\ee
gives the effective action $\Gamma$ as
\be
\Gamma[\phi_c]=\frac{4A}{(1+8A)}\tilde\phi_c^2+\frac{8A}{3(1+8A)^4}(-3+12A+128A^3)\tilde\phi_c^4+\cdots~.
\ee
Notice that this expression coincides with the effective action for 
the $O(N)$-symmetric model (Eq.~\ref{ONgamma}) , when $N=1$ 
(taking into account the difference on the coupling $\lambda$ definition). 
At the large volume limit, we arrive at a form independent of the coupling: 
\be
\Gamma[\phi_c]\simeq \frac{1}{2}\left(\frac{\phi_c}{v} \right)^2+\frac{1}{12} 
\left(\frac{\phi_c}{v} \right)^4+\cdots~.
\ee
Finally, the effective potential is obtained after dividing by the volume $V$, and becomes flat in the limit $V\to\infty$.

\end{document}